\documentclass[a4paper]{spie}  

\usepackage{amsmath,amsfonts,amssymb}
\usepackage{graphicx}
\usepackage[utf8]{inputenc}
\usepackage[colorlinks=true, allcolors=blue]{hyperref}
\usepackage[T1]{fontenc}
\usepackage{textcomp}
\usepackage{newunicodechar}
\usepackage{float}
\usepackage{subcaption}
\usepackage[table]{xcolor}
\usepackage{booktabs}
\usepackage{tabularx}
\newunicodechar{̀}{\`{}}
\definecolor{lightred}{RGB}{255,220,220}

\title{The CUSP CubeSat mission for space weather multi-physics analysis, design, and testing}

\author[a]{Giovanni Lombardi}
\author[g]{Riccardo Di Spirito}
\author[a]{Sergio Fabiani}
\author[a]{Daniele Censi}
\author[b]{Giovanni De Cesare}
\author[a]{Ettore Del Monte}
\author[b,r]{Riccardo Campana}
\author[a]{Enrico Costa}
\author[c]{Mauro Centrone}
\author[a]{Nicolas De Angelis}
\author[a]{Sergio Di Cosimo}
\author[a]{Giuseppe Di Persio}
\author[a]{Abhay Kumar}
\author[a]{Pasqualino Loffredo}
\author[e]{Gabriele Minervini}
\author[a]{Fabio Muleri}
\author[d]{Paolo Romano}
\author[a]{Alda Rubini}
\author[a]{Emanuele Scalise}
\author[a]{Paolo Soffitta}
\author[f,a]{Andrea Alimenti}
\author[f,a]{Enrico Silva}
\author[g]{Bruna Bertucci}
\author[h]{Giovanni Ambrosi}
\author[i]{Giovanni CUcinella}
\author[i]{Vito Di Bari}
\author[i]{Simone Di Filippo}
\author[i]{Andrea Negri}
\author[i]{Massimo Perelli}
\author[l,m]{Dario Modenini}
\author[l]{Andrea Curatolo}
\author[m]{Nicolas Gagliardi}
\author[l]{Daniele Pecorella}
\author[m]{Alice Ponti}
\author[l,m]{Paolo Tortora}
\author[n]{Andrea Del Re}
\author[n]{Davide Albanesi}
\author[n]{Valerio Campamaggiore}
\author[n]{Giulia De Iulis}
\author[n]{Adam Ebrahim}
\author[n]{Marco Schepis}
\author[n]{Alessandro Zambardi}
\author[o]{Ilaria Baffo}
\author[o]{Pierluigi Fanelli}
\author[p]{Costantino Zazza}
\author[q]{Valerio Vagelli}
\author[q]{Daniele Brienza}
\author[q]{Immacolata Donnarumma}
\author[q]{Matteo Mergè}
\author[q]{Emanuele Zaccagnino}
\author[q]{Alessandro Turchi}

\affil[a]{INAF-IAPS, via del Fosso del Cavaliere 100, 00133 Rome, Italy}
\affil[b]{INAF-OAS Bologna, via Gobetti 93/3, 40129, Bologna, Italy}
\affil[c]{INAF-OAR, Via Frascati 33, 00040, Monte Porzio Catone, Italy}
\affil[d]{INAF-OACT, Via S. Sofia 78, 95123, Catania, Italy}
\affil[e]{INAF-Headquarters, Viale del Parco Mellini 84, 00136, Rome, Italy}
\affil[f]{Department of Industrial, Electronic and Mechanical Engineering, Roma Tre University, via V. Volterra 62, 00146, Rome, Italy}
\affil[g]{Department of Physics and Geology, University of Perugia, Via Alessandro Pascoli snc, 06123, Perugia, Italy}
\affil[h]{INFN Sezione di Perugia, Via Alessandro pascoli snc, 06123, Perugia, Italy}
\affil[i]{IMT s.r.l., via Carlo Bartolomeo Piazza 30, 00161, Rome, Italy}
\affil[l]{Interdepartmental Centre for Industrial Aerospace Research - Alma Mater Studiorum Università di Bologna - Via Carnaccini 12, 47121 Forlı̀, Italy}
\affil[m]{Department of Industrial Engineering - Alma Mater Studiorum Università di Bologna - Via Montaspro 97, 47121 Forlı̀, Italy}
\affil[n]{DEDA Connect s.r.l., Via Vincenzo Lamaro 51, 00173, Rome, Italy}
\affil[o]{DEIM, University of Tuscia, Largo dell’Università, 01100, Viterbo, Italy}
\affil[p]{DIBAF, University of Tuscia, Largo dell’Università, 01100, Viterbo, Italy}
\affil[q]{ASI, Via del Politecnico snc, 00133, Rome, Italy}
\affil[r]{INFN Sezione di Bologna, viale Berti Pichat 6/2,  40127, Bologna, Italy}

\authorinfo{Further author information: (Send correspondence to G. Lombardi)\\G. Lombardi: E-mail: giovanni.lombardi@inaf.it}

\begin{document} 
\maketitle
\newpage
\begin{abstract}
The CUbesat Solar Polarimeter (CUSP) mission aims to measure the linear polarization of solar flares in the hard X-ray band by means of a Compton scattering polarimeter. CUSP is a project in the framework of the Alcor Program of the Italian Space Agency aimed at developing new CubeSat missions. We present the outcomes of the CUSP's Phase B study, which is ended on 2 July 2026.
The design solutions adopted for the mission's most critical multi-physics design drivers will be discussed, these solutions have been formulated and applied to demonstrate compliance with system requirements at both the spacecraft and platform levels. Moreover, we will discuss the validation of the Payload model based on the environmental testing campaign (e.g., vibration) carried out on a demonstrator.  
\end{abstract}
\keywords{Mechanical, Vibration, FEM, Correlation, Optimization}

\section{INTRODUCTION}
\label{sec:intro}  
CubeSat platforms have become increasingly attractive for scientific space missions owing to their reduced development time, lower costs, and high flexibility \cite{ECSS2022,NASA2021}. Despite their compact size, they are now capable of hosting advanced scientific payloads, introducing significant challenges in terms of structural design, mechanical integration, and environmental qualification.\\
The CubeSat Solar Polarimeter (CUSP) is a 6U XL CubeSat mission developed by a consortium of public institutions and Industries lead by the Italian National Institute for Astrophysics (INAF)\cite{Fabiani2026} to investigate the relationship between solar energetic events and space weather\cite{Fabiani2022,Fabiani2024}. The payload integrates scintillation detectors, photodetectors, front-end and back-end electronics, and a dedicated collimator assembly within the limited volume available on a CubeSat platform, requiring a lightweight yet mechanically robust design capable of withstanding the launch environment.\\
To verify the structural integrity of the instrument, a representative Structural Model (SM) was developed and analysed through finite element modelling in accordance with ECSS and NASA GEVS requirements. The numerical predictions were subsequently validated through a qualification-level vibration test campaign performed along the three orthogonal axes.\\
This paper presents the structural design of the CUSP payload, the development of the finite element model, the environmental vibration test campaign, and the correlation between numerical and experimental results, demonstrating the capability of the adopted modelling approach to accurately predict the dynamic behaviour of the instrument and providing a validated structural baseline for the subsequent development phases.

\section{Payload Design and Structural Architecture}
\subsection{Payload Architecture}
The CUSP payload has been developed for integration within a 6U XL CubeSat platform and consists of two main units: the Front-End Unit (FEU) and the Back-End Unit (BEU). The FEU hosts the scientific detector system, including the scintillator assemblies, photodetectors, front-end electronics, and the collimator assembly, while the BEU accommodates four Back-End Electronics (BEE) boards responsible for power distribution, data handling, and instrument control\cite{Lombardi2024,Lombardi2025}.\\
The payload has been designed to satisfy the stringent mass, stiffness, and envelope constraints imposed by the CubeSat platform while ensuring mechanical robustness under launch loads. The mechanical architecture, Figure~1, combines lightweight aluminium alloy structural components with dedicated detector supports, providing accurate alignment of the sensitive elements and efficient load transfer toward the spacecraft interfaces.
\begin{figure}[H]
    \centering
    \begin{subfigure}{0.45\textwidth}
        \centering
        \includegraphics[width=\linewidth]{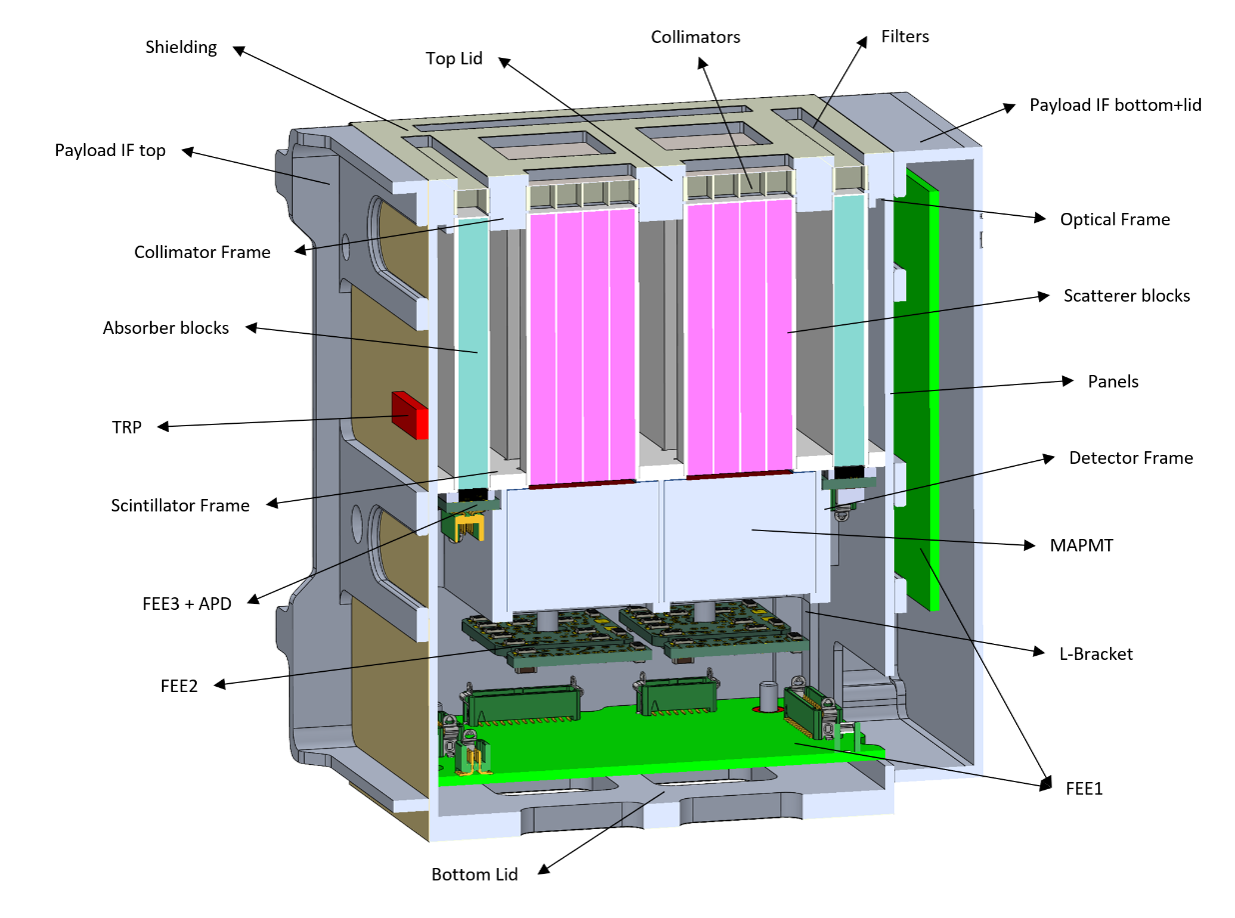}
        \caption{Front-End Unit}
        \label{fig:feu}
    \end{subfigure}
    \hfill
    \begin{subfigure}{0.4\textwidth}
        \centering
        \includegraphics[width=\linewidth]{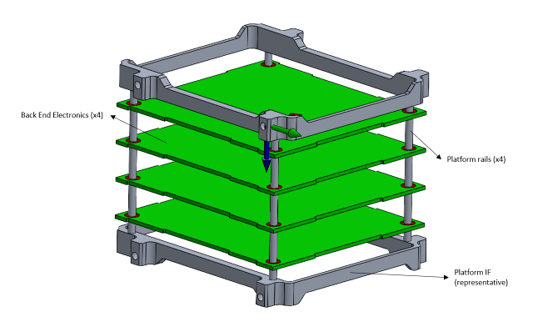}
        \caption{Back-End Unit}
        \label{fig:beu}
    \end{subfigure}
    \caption{CUSP payload architecture.}
    \label{fig:payload_architecture}
\end{figure}

\subsection{Structural Model}
A representative Structural Model (SM) was developed to verify the mechanical integrity of the payload under launch conditions. The Structural Model (SM) is representative of the Front End Unit that is the most critical custom designed part of the Payload, the model reproduces the flight mechanical architecture and includes representative masses and interfaces together with representative critical components, namely one Multianode PhotoMultipler Tubes (MAPMT), one board equipped with eight Avalanche Photodiodes (APD), one plastic scintillator assembly, one inorganic scintillator assembly, and two tungsten collimators. The remaining non-critical elements are represented by dedicated dummy components to preserve the overall mass and stiffness distribution.\\
The Structural Model in Figure~2 was used both for finite element correlation and for the qualification-level vibration test campaign performed along the three orthogonal axes\cite{Lombardi2025}.
\begin{figure}[H]
    \centering
    \begin{subfigure}{0.3\textwidth}
        \centering
        \includegraphics[width=\linewidth]{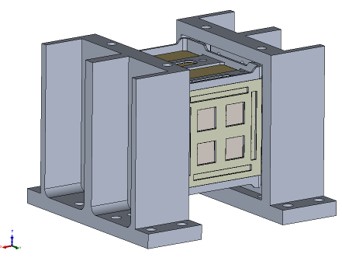}
        \caption{Structural Model (CAD)}
        \label{fig:feu}
    \end{subfigure}
    \hfill
    \begin{subfigure}{0.3\textwidth}
        \centering
        \includegraphics[width=\linewidth]{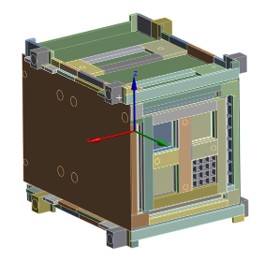}
        \caption{Structural Model (FEM)}
        \label{fig:beu}
    \end{subfigure}
    \caption{CUSP structural model.}
    \label{fig:payload_architecture}
\end{figure}

\section{Finite Element Model Development}
\subsection{Finite Element Model}
A detailed finite element model of the CUSP Structural Model was developed in ANSYS Mechanical to support the structural verification activities and predict the payload response under launch environments. The model reproduces the global stiffness, mass distribution, and mechanical interfaces of the flight configuration while preserving an efficient computational cost suitable for iterative design activities.\\
The FE model includes the primary aluminium structure, detector supports, scintillator assemblies, collimator supports, interface brackets, and representative electronic boards. Particular attention was devoted to reproducing the mechanical interfaces and local stiffness of the detector assembly, as these represent the most critical contributors to the payload dynamic response.\\
The model was developed using a combination of solid and shell finite elements with local mesh refinement at structural interfaces and load transfer regions, Figure~3. Material properties and interface definitions were implemented according to the current mechanical design and verified engineering data.
   \begin{figure} [H]
   \begin{center}
   \begin{tabular}{c} 
   \includegraphics[height=5cm]{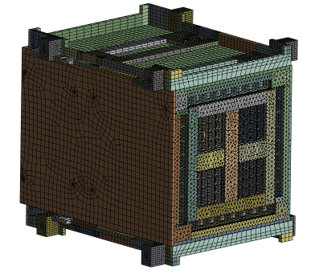}
	\end{tabular}
	\end{center}
   \caption{CUSP FEU Mesh}
   { \label{DA}}
 \end{figure} 
The resulting model was specifically developed to preserve the global stiffness characteristics and dynamic behaviour of the payload while maintaining an efficient computational cost suitable for iterative design activities, as shown in Figure~4. The FEM model also reproduces the mass of the real structural model, with a total mass of approximately 1,052 g, thereby improving the fidelity of the numerical representation.
   \begin{figure}[H]
   \begin{center}
   \begin{tabular}{c} 
   \includegraphics[height=5cm]{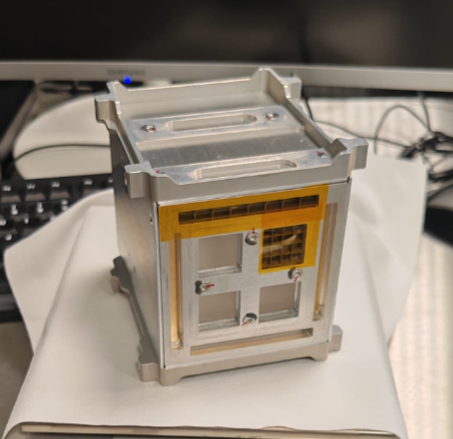}
	\end{tabular}
	\end{center}
   \caption{Structural Model}
   { \label{fig:feu}}
 \end{figure} 
\subsection{Verification Approach}
The structural verification campaign combined numerical simulations and experimental testing to demonstrate compliance with the launch environment requirements. Modal analyses were performed to assess the payload dynamic behaviour and verify launcher compatibility, while quasi-static and random vibration analyses were used to evaluate structural integrity under qualification loading conditions.\\
The numerical analyses were performed in accordance with ECSS standards and NASA General Environmental Verification Standard (GEVS) requirements\cite{ECSS2022,NASA2021}, adopting the same boundary conditions and mechanical interfaces later implemented during the experimental vibration campaign. This approach ensured direct correlation between numerical predictions and experimental measurements.

\subsection{Numerical Results}
In Table~1 and Figure~5, the finite element analyses confirmed the structural compliance of the payload with the project requirements. The first significant structural mode was predicted at approximately 595 Hz, well above the minimum launcher compatibility requirement of 120 Hz, demonstrating the high stiffness of the adopted architecture.\\
Quasi-static analyses confirmed positive margins of safety for all primary structural components, while the random vibration simulations identified the dominant resonance frequencies and dynamic response of the payload. These results constituted the reference baseline for the subsequent vibration test campaign and the FEM-to-test correlation activities discussed in Section~5.
\begin{table*}[ht]
\centering
\caption{Modal analysis results and effective mass participation factors.}
\label{tab:modal_results}
\scriptsize
\begin{tabular}{cccccccc}
\hline
Mode & Frequency & Eff. Mass &
Eff. Mass &
Eff. Mass &
Eff. Mass &
Eff. Mass &
Eff. Mass \\
No. & [Hz] & UX [\%] & UY [\%] & UZ [\%] & ROTX [\%] & ROTY [\%] & ROTZ [\%] \\
\hline
1  & 594.97 & 1.63 & 0.00 & 0.00 & 0.01 & 0.07 & 1.13 \\
2  & 596.06 & 0.29 & 0.00 & 0.00 & 0.00 & 0.39 & 0.20 \\
3  & 624.20 & 0.02 & 0.00 & 0.00 & 1.21 & 0.00 & 0.01 \\
4  & \cellcolor{lightred}741.63 & 0.00 & 4.03 & 94.56 & 0.00 & 0.00 & 0.00 \\
5  & 883.02 & 0.00 & 1.68 & 0.05 & 0.01 & 0.00 & 0.00 \\
6  & 1189.50 & 0.79 & 0.00 & 0.00 & 3.20 & 0.00 & 0.50 \\
7  & \cellcolor{lightred}1192.40 & 4.22 & 0.00 & 0.00 & 0.30 & 0.00 & 2.59 \\
8  & \cellcolor{lightred}1245.00 & 62.02 & 0.03 & 0.01 & 0.24 & 0.00 & 0.56 \\
9  & 1245.80 & 0.19 & 0.00 & 0.00 & 0.00 & 0.00 & 0.00 \\
10 & \cellcolor{lightred}1398.00 & 0.37 & 6.77 & 0.00 & 49.20 & 0.01 & 3.19 \\
11 & \cellcolor{lightred}1430.00 & 0.00 & 44.83 & 0.01 & 8.28 & 1.01 & 1.79 \\
12 & 1502.10 & 0.96 & 3.16 & 0.00 & 0.66 & 1.38 & 61.18 \\
13 & 1581.30 & 0.00 & 1.39 & 0.01 & 0.01 & 64.67 & 0.97 \\
14 & 1621.00 & 0.01 & 0.00 & 0.00 & 0.23 & 0.00 & 0.01 \\
15 & 1626.40 & 0.00 & 0.25 & 0.02 & 0.58 & 2.57 & 0.12 \\
16 & 1636.80 & 0.00 & 0.44 & 0.08 & 0.02 & 0.03 & 0.00 \\
17 & 1637.00 & 0.00 & 0.01 & 0.01 & 0.61 & 0.01 & 0.00 \\
18 & 1655.00 & 0.09 & 0.11 & 0.36 & 0.39 & 0.02 & 0.09 \\
19 & 1678.30 & 0.00 & 0.06 & 0.00 & 0.12 & 0.07 & 0.22 \\
20 & 1684.50 & 0.04 & 0.79 & 0.01 & 0.00 & 0.00 & 0.16 \\
21 & 1694.70 & 0.21 & 0.49 & 0.01 & 0.07 & 0.04 & 0.48 \\
22 & 1703.70 & 0.03 & 3.00 & 0.00 & 0.01 & 0.55 & 0.01 \\
23 & 1745.50 & 0.76 & 0.20 & 0.00 & 0.08 & 0.01 & 0.04 \\
24 & 1760.80 & 0.00 & 0.06 & 0.00 & 0.00 & 2.03 & 0.01 \\
25 & 1762.70 & 0.13 & 1.67 & 0.01 & 0.02 & 0.04 & 0.10 \\
26 & \cellcolor{lightred}1826.30 & 0.57 & 9.07 & 0.00 & 0.06 & 0.17 & 0.02 \\
27 & 1950.80 & 0.47 & 0.19 & 0.05 & 14.05 & 0.00 & 0.07 \\
\hline
\end{tabular}
\end{table*}
\begin{figure} [H]
   \begin{center}
   \begin{tabular}{c} 
   \includegraphics[height=7.5cm]{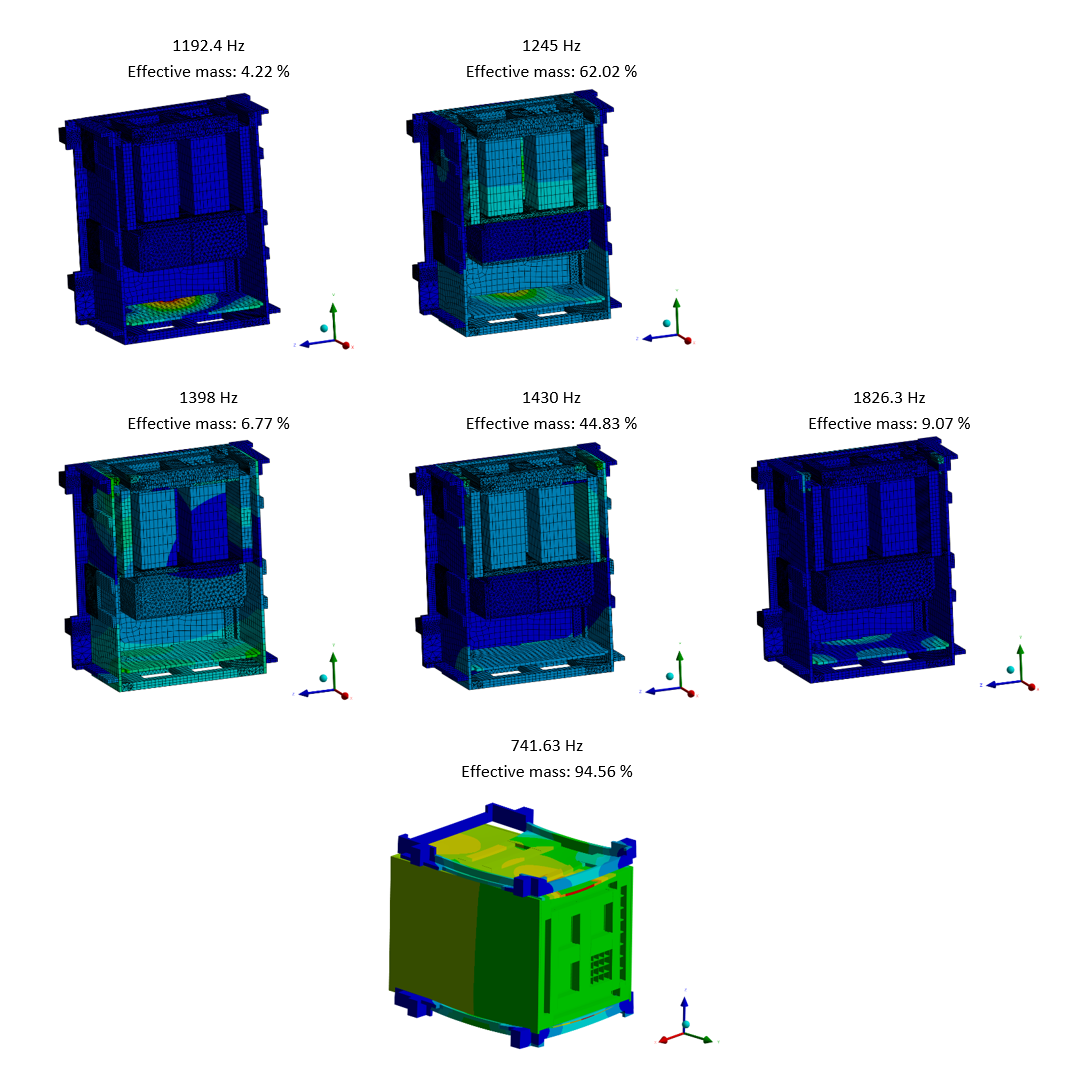}
	\end{tabular}
	\end{center}
   \caption{PL high-mass modes for UZ, UY and UZ}
   { \label{DA}}
 \end{figure}
 
\section{Environmental Vibration Test}
The representative Structural Model described in Section 2 was subjected to a qualification-level vibration campaign to experimentally verify the structural integrity of the payload and validate the finite element model.\\
Unlike conventional mass simulators, the test article incorporated several flight-representative components, including one MAPMT, one baord equipped with eight APDs, one plastic scintillator assembly, one inorganic scintillator assembly, and two tungsten collimators. This configuration enabled both structural verification and post-test functional inspections of the most mechanically sensitive subsystems.
\subsection{Test Setup}
The vibration campaign was carried out on 13--14 April 2026 at the SERMS laboratory (Terni, Italy) using an electrodynamic shaker facility qualified for spacecraft environmental testing, Figure~6.\\
The Structural Model was mounted on a dedicated aluminium fixture reproducing the flight mechanical interface with the CubeSat platform. The specimen was instrumented, as shown in Table~2, with control and response accelerometers to monitor both the input excitation and the payload dynamic response throughout the test campaign.\\
For each excitation axis, the test sequence consisted of a pre-random resonance search, a qualification-level random vibration exposure, and a post-random resonance search to assess possible variations in the dynamic behaviour of the payload.
\begin{figure}[H]
    \centering
    \begin{subfigure}{0.5\textwidth}
        \centering
        \includegraphics[width=\linewidth]{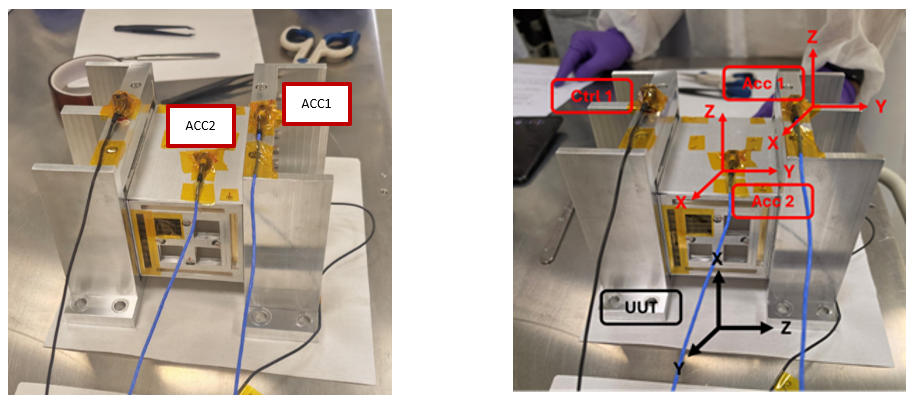}
        \caption{X-axis}
        \label{fig:X}
    \end{subfigure}
    \hfill
    \begin{subfigure}{0.5\textwidth}
        \centering
        \includegraphics[width=\linewidth]{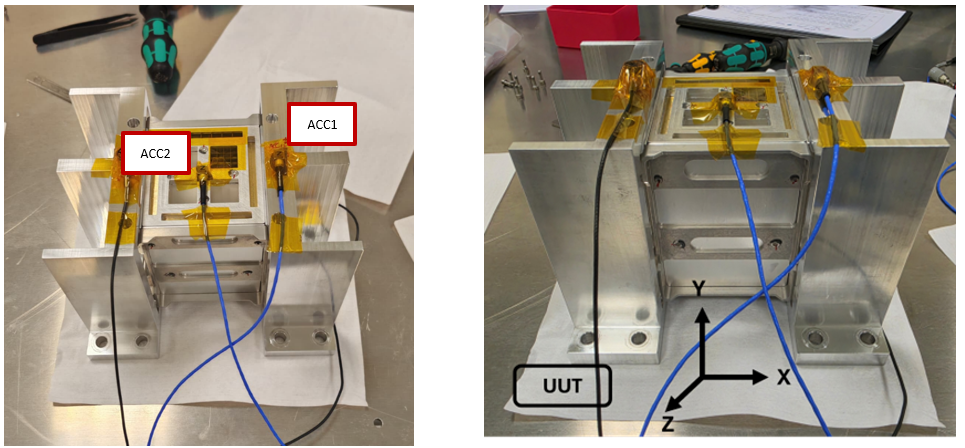}
        \caption{Y-axis}
        \label{fig:Y}
    \end{subfigure}
    \hfill
    \begin{subfigure}{0.5\textwidth}
        \centering
        \includegraphics[width=\linewidth]{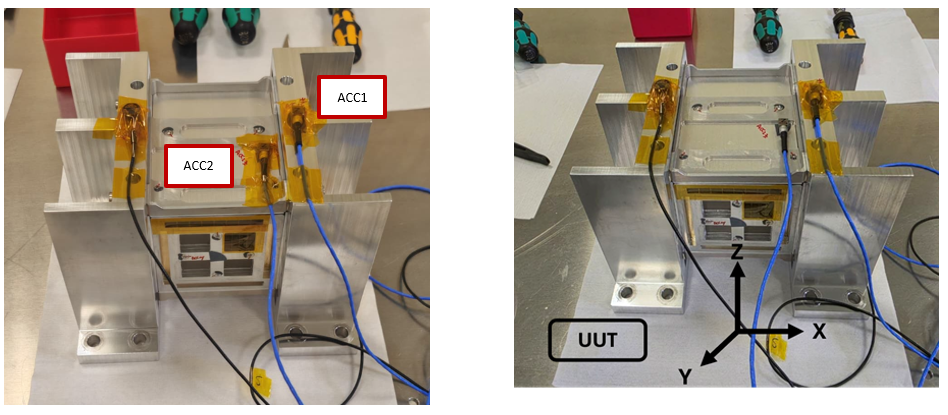}
        \caption{Z-axis}
        \label{fig:Z}
    \end{subfigure}
    \caption{Setup of the test.}
    \label{fig:payload_architecture}
\end{figure}

\begin{table}[ht]
\centering
\caption{Accelerometer configuration adopted during the vibration test campaign.}
\label{tab:accelerometers}
\begin{tabularx}{\linewidth}{l l l l X}
\toprule
ID & Type & Axis & Position \\
\midrule
Ctrl 1 & Control & Mono & Top of Bracket (left) \\
Ctrl 2 & Control & Mono & Bottom of Bracket (right) \\
Acc 1 & Monitor & X, Y, Z & Top of Bracket (left) \\
Acc 2 & Measure & X, Y, Z & See Figure 8 \\
\bottomrule
\end{tabularx}
\end{table}
\subsection{Test Conditions}
The environmental verification campaign was performed according to the qualification levels defined for the project and in accordance with the adopted verification plan in Table~3. Random vibration tests were executed independently along the three orthogonal payload axes.\\
\begin{table}[H]
\centering
\caption{Qualification vibration test levels.}
\begin{tabular}{lc}
\hline
Test & Level \\
\hline
Resonance Search & 0.5 g \\
Random Vibration & 14.1 g RMS \\
Frequency Range & 20--4000 Hz \\
Duration & 120 s/axis \\
\hline
\end{tabular}
\end{table}
The acquired resonance search data constituted the experimental reference for the FEM correlation presented in Section~5. In addition, visual inspections and functional checks were performed throughout the campaign to verify the integrity of the most critical structural and detector components before and after environmental testing.

\section{Experimental Validation and FEM Correlation}
The qualification-level vibration campaign provided the experimental data required to validate the finite element model and assess the structural behaviour of the CUSP payload under representative launch environments.\\
The comparison between pre- and post-random resonance searches, together with the post-test inspection activities, confirmed that the payload successfully withstood the applied qualification loads without evidence of structural degradation. The acquired experimental data were subsequently used to correlate the finite element predictions with the measured dynamic response.

\subsection{Finite Element Model Correlation}
Figure~7 compares the measured transmissibility functions with the numerical predictions obtained from the finite element model for the three excitation axes.
Overall, the correlation demonstrates very good agreement between numerical and experimental results. The dominant resonance frequencies are accurately reproduced by the numerical model, with discrepancies generally below 100 Hz. Although differences are observed in the absolute transmissibility amplitudes, the numerical model successfully captures the global dynamic behaviour of the payload, confirming the accuracy of the adopted modelling assumptions.
\begin{figure}[H]
    \centering
    \begin{subfigure}{0.7\textwidth}
        \centering
        \includegraphics[width=\linewidth]{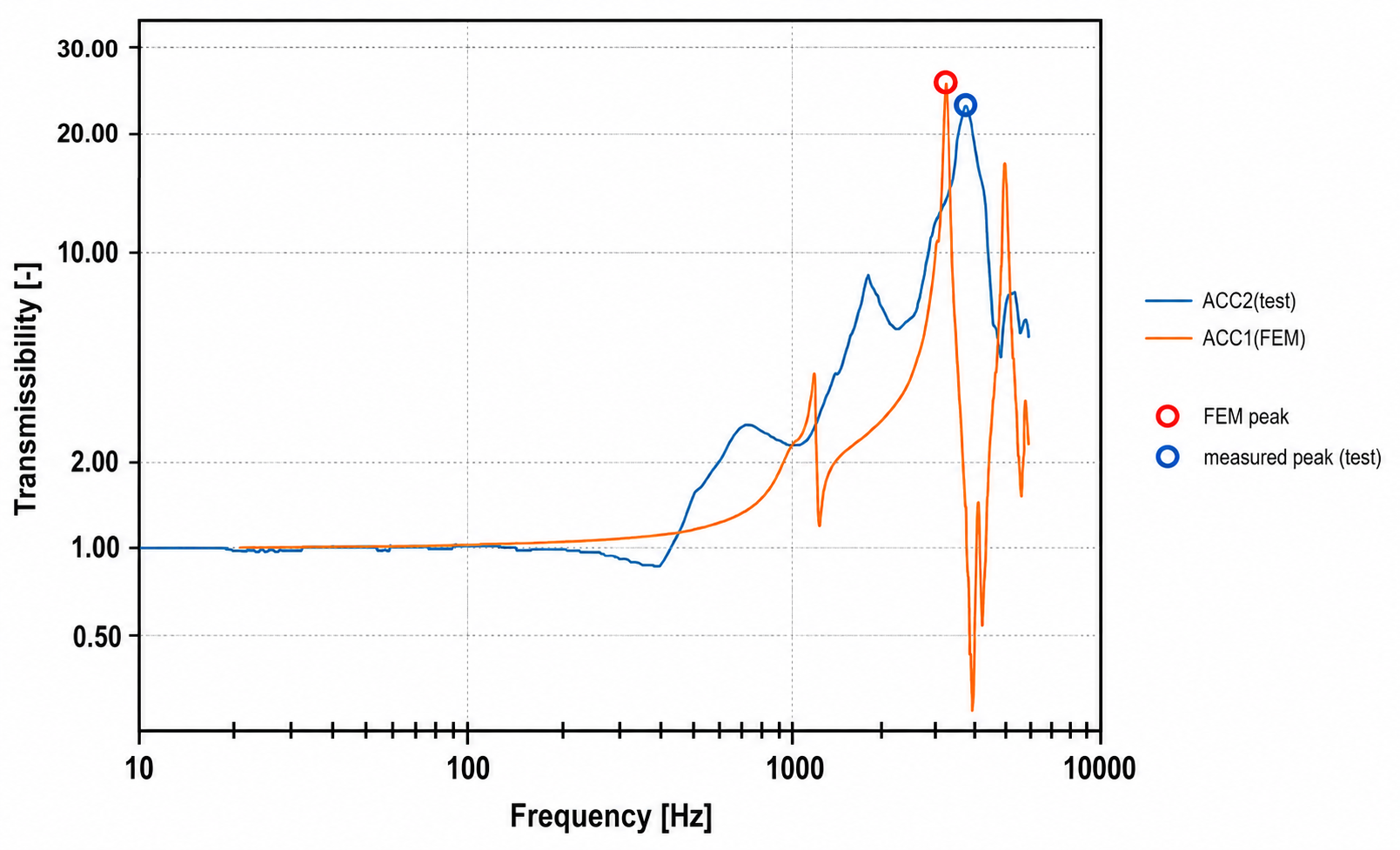}
        \caption{X-axis input}
        \label{fig:X}
    \end{subfigure}
    \hfill
    \begin{subfigure}{0.7\textwidth}
        \centering
        \includegraphics[width=\linewidth]{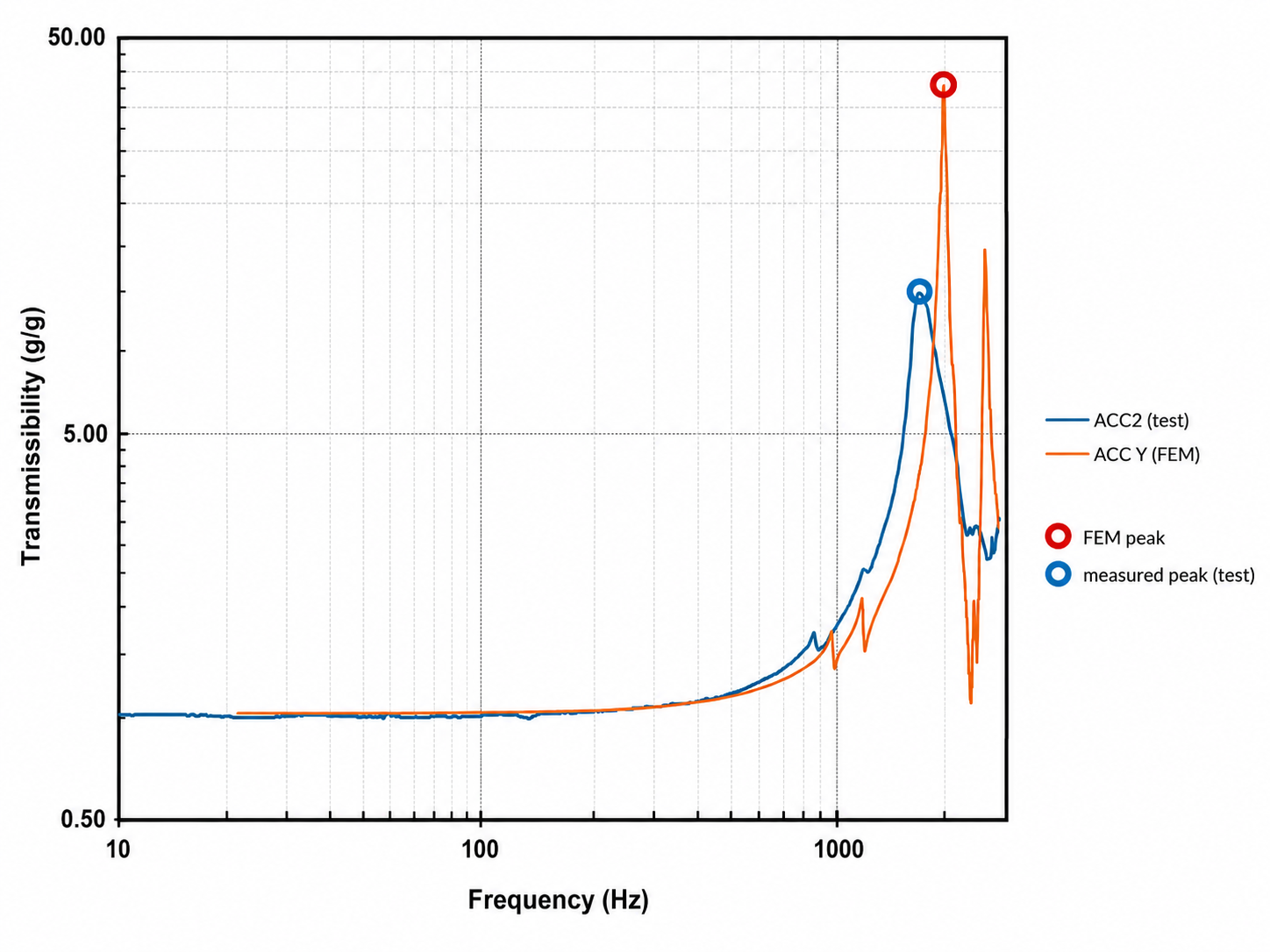}
        \caption{Y-axis input}
        \label{fig:Y}
    \end{subfigure}
    \hfill
    \begin{subfigure}{0.7\textwidth}
        \centering
        \includegraphics[width=\linewidth]{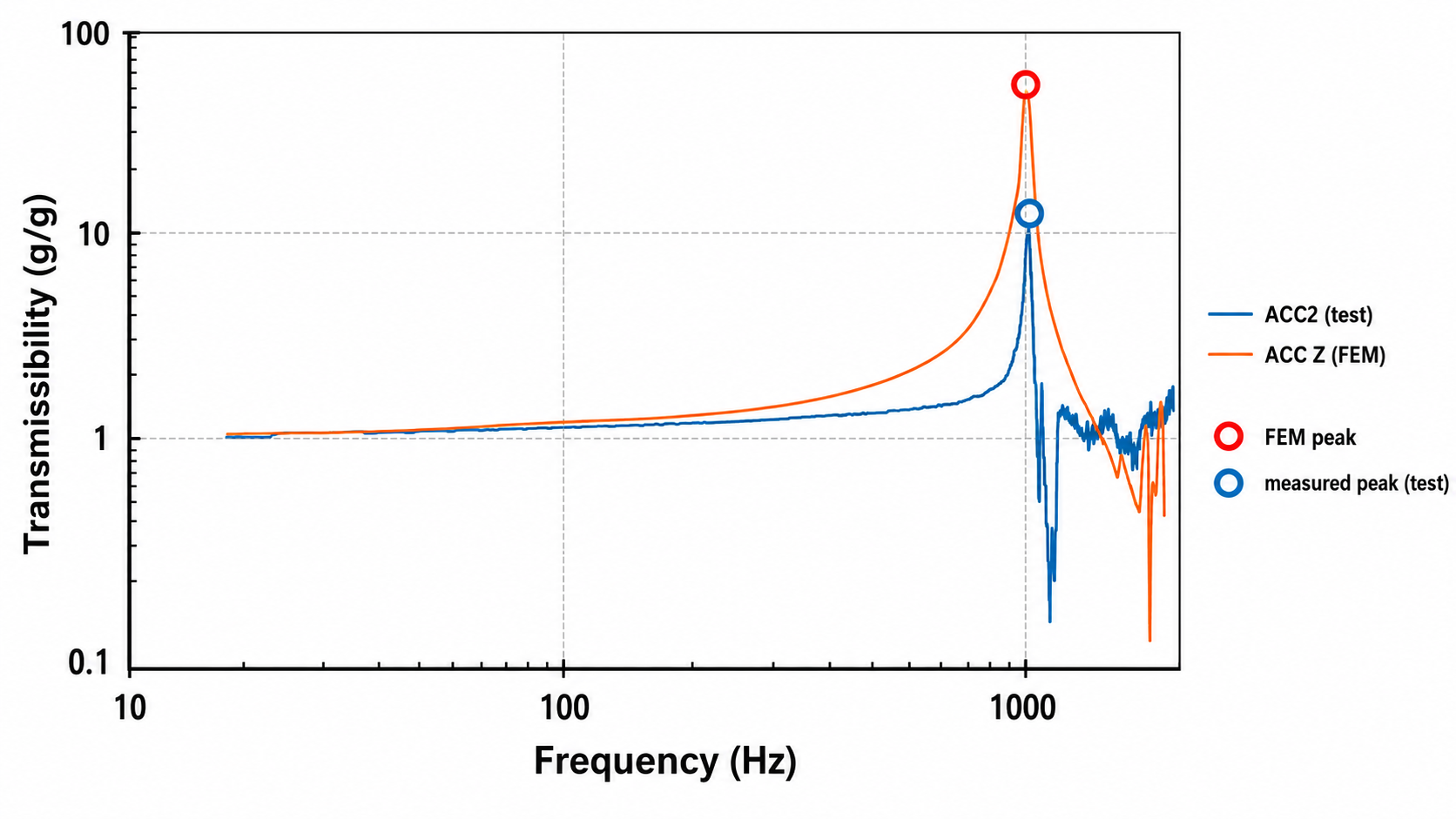}
        \caption{Z-axis input}
        \label{fig:Z}
    \end{subfigure}
    \caption{ACC2 in-axis response.}
    \label{fig:payload_architecture}
\end{figure}

\subsection{Y-axis Dynamic Behaviour}
Unlike the X and Z directions, the Y-axis response exhibited a repeatable transition from multiple resonance peaks to a single dominant peak following random vibration excitation.\\
To investigate this behaviour, three independent vibration sequences were performed. The observed response proved to be fully repeatable, with the initial resonance configuration recovering after the subsequent Z-axis test and reproducing the same evolution during the third Y-axis execution.\\
The repeatability of the phenomenon, in Figure~8, together with the absence of structural damage observed during post-test inspections, indicates that the measured behaviour is an intrinsic characteristic of the payload dynamic response rather than evidence of structural degradation.
\begin{figure} [H]
   \begin{center}
   \begin{tabular}{c} 
   \includegraphics[height=7cm]{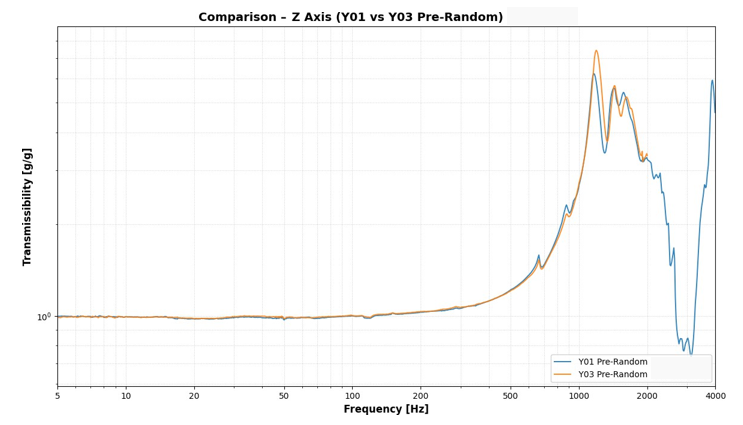}
	\end{tabular}
	\end{center}
   \caption{Comparison between the pre-random sine sweep tests of the first (Y01) and third (Y03) Y-axis executions}
   { \label{ycomparison}}
 \end{figure}
The repeatability of the phenomenon and the absence of any detectable structural anomalies support the conclusion that the payload maintained its mechanical integrity throughout the entire qualification campaign.\\
Visual inspections, microscopic examinations, and functional checks confirmed the absence of permanent deformations, fastener loosening, detector displacement, or degradation of the mechanical interfaces.

\section{Conclusions}
This work presented the structural verification activities performed for the CUSP payload during the Phase B development of the mission, combining numerical modelling and experimental testing to assess the mechanical behaviour of the instrument under representative launch environments.\\
A detailed finite element model was developed to support the structural design process and predict the payload response under modal, quasi-static, and random vibration loading. The numerical activities were complemented by a qualification-level vibration test campaign performed on a flight-representative Structural Model, specifically developed to reproduce the mass, stiffness, and mechanical interfaces of the flight configuration while incorporating the most critical detector components.\\
The comparison between numerical predictions and experimental measurements demonstrated good agreement in the identification of the dominant structural resonances, validating the adopted modelling approach and confirming the capability of the finite element model to accurately reproduce the global dynamic behaviour of the payload. The environmental campaign further demonstrated the robustness of the mechanical design, with post-test inspections confirming the absence of structural failures, permanent deformations, or degradation of the critical interfaces and detector assemblies.\\
The activities presented in this work establish a validated structural baseline for the CUSP payload and provide confidence in the adopted design methodology. The correlated finite element model and the experimental results obtained during the vibration campaign will support the refinement of the payload design and the development of the Engineering Qualification Model (EQM), contributing to the subsequent qualification and flight preparation phases of the CUSP mission.

\acknowledgments 
 This work is funded by the Italian Space Agency (ASI) within the Alcor Program, as part of the development of the CUSP mission under contract n. 2023-2-R.0.

\bibliography{report} 
\bibliographystyle{spiebib} 

\end{document}